\documentclass[cits]{PoS}

\def\lsim{\lower.5ex\hbox{$\; \buildrel < \over \sim \;$}}
\def\gsim{\lower.5ex\hbox{$\; \buildrel > \over \sim \;$}}

\title{Cosmic-ray acceleration in supernova shocks}

\ShortTitle{Cosmic-ray acceleration in supernova shocks}

\author{\speaker{Vincent Tatischeff}\thanks{Permanent adress: CSNSM, IN2P3-CNRS 
and Univ Paris-Sud, F-91405 Orsay Cedex, France}\\
        Institut de Ci\`encies de l'Espai (CSIC-IEEC), Campus UAB, Fac. Ci\`encies, 
	08193 Bellaterra, Barcelona, Spain\\
        E-mail: \email{Vincent.Tatischeff@csnsm.in2p3.fr}}

\abstract{Galactic cosmic rays are widely believed to be accelerated in expanding 
shock waves initiated by supernova explosions. The theory of diffusive shock 
acceleration of cosmic rays is now well established, but two fundamental questions 
remain partly unanswered: what is the acceleration efficiency, i.e. the fraction of 
the total supernova energy converted to cosmic-ray energy, and what is the maximum 
kinetic energy achieved by particles accelerated in supernova explosions? Recent 
observations of supernova remnants, in X-rays with the {\it Chandra} and 
{\it XMM-Newton} satellites and in very-high-energy $\gamma$ rays with several ground-based
atmospheric Cerenkov telescopes, have provided new pieces of information concerning
these two questions. After a review of these observations and their current
interpretations, I show that complementary information on the diffusive shock 
acceleration process can be obtained by studying the radio emission from extragalactic 
supernovae. As an illustration, a nonlinear model of diffusive shock acceleration is 
applied to the radio light curves of the supernova SN~1993J, which exploded in the
nearby galaxy M81. The results of the model suggest that most of the Galactic cosmic 
rays may be accelerated during the early phase of interaction between the supernova ejecta
and the wind lost from the progenitor star.}

\FullConference{Supernovae: lights in the darkness (XXIII Trobades Cient\'ifiques de la 
                Mediterr\`ania)\\
		 October 3-5 2007\\
		 Mao, Menorca, Spain}

\begin{document}

\section{Introduction}

Gamma-ray observations of the Small Magellanic Cloud with the EGRET telescope onboard 
the {\it Compton Gamma Ray Observatory} have proved that the bulk of cosmic rays (CRs) 
propagating in the Milky Way are produced in Galactic sources \cite{sre93}. 
Observations of the diffuse $\gamma$-ray emission from our Galaxy allow to estimate 
the total CR luminosity \cite{dog02}:
\begin{equation}
L_{\rm CR} = L_\gamma {x_\gamma \over x} \sim 5 \times 10^{40}~{\rm erg~s}^{-1},
\label{eqvt1}
\end{equation}
where $L_\gamma \sim 5 \times 10^{39}$~erg~s$^{-1}$ is the total 
luminosity of diffuse high-energy ($>100$~MeV) $\gamma$ rays emitted in the decay of 
$\pi^0$ produced by CR interaction with the interstellar medium (ISM), $x_\gamma 
\sim 120$~g~cm$^{-2}$ is the mean grammage needed for a CR ion to produce 
a $\pi^0$ in the ISM and $x \sim 12$~g~cm$^{-2}$ is the mean path length that CRs 
traverse before escaping the Galaxy, which is determined from measurements of the CR
chemical composition near Earth. In comparison, the total power supplied by Galactic 
supernovae (SNe) is
\begin{equation}
L_{\rm SN} = E_{\rm SN} R_{\rm SN} \approx 10^{42}~{\rm erg~s}^{-1},
\label{eqvt2}
\end{equation}
where $E_{\rm SN} \approx 1.5\times 10^{51}$~erg is the approximate total ejecta kinetic 
energy of a SN and $R_{\rm SN} \approx 2$ per century is the current epoch 
Galactic SN rate \cite{fer98}. Thus, SNe have enough power to sustain the CR 
population against escape from the Galaxy and energy losses, if there is a mechanism 
for channeling $\sim 5$\% of the SN mechanical energy release into relativistic 
particles.

Diffusive shock acceleration (DSA) at the blast waves generated by SN explosions can 
in principle produce the required acceleration efficiency, as well as the observed 
power-law spectrum of CRs \cite{kry77,axf78,bel78,bla78}. In this model, a fraction 
of ambient particles entering the SN shock front can be accelerated to high energies 
during the lifetime of a supernova remnant (SNR) by diffusing back and forth on 
compressive magnetic fluctuations of the plasma flow on both sides of the shock. A 
critical ingredient of the theory is the strength of the turbulent magnetic field 
in the shock acceleration region, which governs the acceleration rate and in turn 
the maximum energy of the accelerated particles. If the turbulent field upstream of 
the SN shock is similar to the preexisting field in the surrounding ISM ($B \sim
5$~$\mu$G), the maximum total energy of an ion of charge $Z$ was estimated 25 years ago 
to be (for a quasi-parallel shock geometry) $E_{\rm max} \sim 10^{14} Z$~eV \cite{lag83}. 
But in more recent developments of the DSA theory, it is predicted that large-amplitude 
magnetic turbulence is self-generated by streaming of accelerated particles in the 
shock region, such that the ambient magnetic field can be strongly amplified as part 
of the acceleration process \cite{bel01,ama06,vla06}. In this case, protons might be 
accelerated in SNRs up to $3\times10^{15}$~eV, i.e. the energy of the spectral "knee"   
above which the measured all-particle CR spectrum shows a significant steepening. 
Contributions of accelerated $\alpha$-particles and heavier species might then explain 
the existing CR measurements up to $\sim$10$^{17}$~eV \cite{ber07}. Above energies of 
10$^{18}$--10$^{19}$~eV, CRs are probably of extragalactic origin.  

Another uncertain parameter of the DSA model is the fraction of total shocked 
particles injected into the acceleration process. Although theoretical progress has 
been made in recent years \cite{bla05}, the particle injection and consequently the 
acceleration efficiency are still not well known. However, theory predicts that for 
efficient acceleration the energy density of the relativistic nuclear component can 
become comparable to that of the postshock thermal component, in which case the 
backreaction of energetic ions can significantly modify the shock structure and 
the acceleration process can become highly nonlinear (e.g. \cite{ber99}). In 
particular, the compression ratio of a CR-modified shock is expected to be higher 
than for a test-particle shock (i.e. when the accelerated particles have no 
influence on the shock structure). 
This is because of both the softer equation of state of a relativistic (CR) gas and 
the energy loss due to escape of accelerated particles from the shock region 
\cite{dec00}. Moreover, the temperature of the shock-heated gas can be reduced if a 
significant fraction of the total available energy of the shock goes into relativistic 
particles. Observations of these nonlinear effects \cite{hug00,dec05,war05} provide 
indirect evidence for the efficient acceleration of ions in SN shock waves. 

The acceleration of electrons in SNRs leaves no doubt, since we observe the nonthermal 
synchrotron emission that these particles produce in the local magnetic field. Radio 
synchrotron radiation, which in SNRs is emitted by GeV electrons, was discovered 
in the 1950's. More recent is the observation of X-ray synchrotron emission from 
young shell-type SNRs \cite{koy95}, which is due to electrons accelerated to very high 
energies, $E_e>$1~TeV. Thanks to the extraordinary spectroscopic-imaging capabilities of 
the {\it XMM-Newton} and {\it Chandra} X-ray observatories, this nonthermal emission can 
now be studied in great details and recent observations of SNRs with these satellites 
have shed new light on the DSA rate and the maximum energy of the accelerated particles. 
This is the subject of Section~2. 

In Section~3, we discuss the origin of the TeV $\gamma$-ray emission observed from 
a handful of shell-type SNRs with atmospheric Cerenkov telescopes. For some objects, 
the detected $\gamma$-rays have been explained as resulting from $\pi^0$ production in 
nuclear collisions of accelerated ions with the ambient gas. If this were true, this 
high-energy emission would be the first observational proof that CR ions are indeed 
accelerated in SN shock waves. However, the origin of the TeV $\gamma$-rays 
emitted in SNRs is still a matter of debate, because at least in some cases the 
high-energy photons can also be produced by inverse Compton scattering of 
cosmic-microwave-background photons (and possibly optical and infrared interstellar 
photons) by ultrarelativistic electrons. 

In Section 4, we show that radio observations of extragalactic SNe can 
provide complementary information on the DSA mechanism. As an example, we use a 
semianalytic description of nonlinear DSA to model the radio light curves 
of SN 1993J. We choose this object because the set of radio data 
accumulated over the years \cite{wei07} constitutes one of the most detailed sets of 
measurements ever established for an extragalactic SN in any wavelength range. We 
derive from these data constraints on the magnetic field strength in the environment 
of the expanding SN shock wave, the maximum energy of the accelerated particles, as 
well as on the fractions of shocked electrons and protons injected into the 
acceleration process. Conclusions are given in Section 5.

\section{X-ray synchrotron emission from SNRs}

Together with the thermal, line-dominated X-ray emission from the shock-heated gas, 
a growing number of SNRs show nonthermal, featureless emission presumably produced
by ultrarelativistic electrons in the blast wave region via a synchrotron process. 
High-angular resolution observations made with the {\it Chandra} and 
{\it XMM-Newton} X-ray observatories have revealed very thin rims of nonthermal 
emission associated with the forward shock. In several cases, 
like SN~1006 \cite{koy95} and G347.3-0.5 \cite{cas04}, the synchrotron component 
completely dominates the thermal X-ray emission. The measured power-law spectral 
index of the X-ray synchrotron radiation is always much steeper than that of the 
nonthermal radio emission, which is consistent with expectation that the X-ray 
domain probes the high-energy end of the accelerated electron distribution. The 
comparison of radio and X-ray fluxes allows to determine the exponential cutoff 
(maximum) frequency of the synchrotron emission, $\nu_c$, which is related to the 
maximum energy of the accelerated electrons and the ambient magnetic field as 
(e.g. \cite{sta06})
\begin{equation}
\nu_c = 1.26\times10^{16} \bigg({E_{e,\rm max} \over 10{\rm~TeV}}\bigg)^2 
\bigg({B \over 10~\mu{\rm G}}\bigg)~{\rm Hz}.
\label{eqvt3}
\end{equation}
Diffusive shock acceleration can only occur for particles whose acceleration rate 
is higher than their energy loss rate in the acceleration region. The maximum 
electron energy, $E_{e,\rm max}$, can be estimated by equating the synchrotron 
cooling time (e.g. \cite{par06}),
\begin{equation}
\tau_{\rm syn}(E_{e,\rm max}) = {E_{e,\rm max} \over (dE/dt)_{\rm syn}} \propto
E_{e,\rm max}^{-1} B^{-2},
\label{eqvt4}
\end{equation}
where $(dE/dt)_{\rm syn}$ is the synchrotron loss rate at $E_{e,\rm max}$, to 
the acceleration time
\begin{equation}
\tau_{\rm acc}(E_{e,\rm max}) = {E_{e,\rm max} \over (dE/dt)_{\rm acc}} \sim
{\kappa(E_{e,\rm max}) \over V_s^2},
\label{eqvt5}
\end{equation}
where $(dE/dt)_{\rm acc}$ and $\kappa(E_{e,\rm max})$ are the acceleration 
rate and mean spatial diffusion coefficient of the electrons of 
energy $E_{e,\rm max}$ in the blast wave region and $V_s$ is the shock speed. 
We have neglected here the dependence of $\tau_{\rm acc}$ on the shock 
compression ratio (see \cite{par06}). The value of $\kappa(E_{e,\rm max})$ 
depends on the strength and structure of the turbulent magnetic field. The DSA 
theory predicts that CRs efficiently excite large amplitude magnetic fluctuations 
upstream of the forward shock and that these fluctuations scatter CRs very 
efficiently \cite{bel78,bel01,ama06,vla06}. It is therefore generally assumed that 
the spatial diffusion coefficient is close to the Bohm limit:
\begin{equation}
\kappa \gsim \kappa_B={r_g v \over 3},
\label{eqvt6}
\end{equation}
where $v$ is the particle speed and $r_g$=$pc/(QeB)$ the particle gyroradius, $p$ 
being the particle momentum, $c$ the speed of light, $Q$ the charge number ($Q=1$ 
for electrons and protons), and $-e$ the electronic charge. Note that for 
ultrarelativistic electrons, $\kappa_B = r_g c/3 \propto E_e B^{-1}$. Equating 
equations~(\ref{eqvt4}) and 
(\ref{eqvt5}) and using equation~(\ref{eqvt3}) to express $E_{e,\rm max}$ as a 
function of $B$ and $\nu_c$, we can write the ratio of the electron diffusion 
coefficient at the maximum electron energy to the Bohm coefficient as:
\begin{equation}
\eta_\kappa={\kappa(E_{e,\rm max}) \over \kappa_B} \propto V_s^2 \nu_c^{-1}.
\label{eqvt7}
\end{equation}
Thus, measurements of $V_s$ and $\nu_c$ can allow to derive $\eta_\kappa$ 
without knowing the ambient magnetic field. Using this result, several 
recent studies \cite{sta06,par06,rey04,yam04} have shown that there are regions 
in young ($t<10^4$~yr) SNRs where {\it acceleration occurs nearly as fast as the 
Bohm theoretical limit} (i.e. $1<\eta_\kappa<10)$. This provides an important 
confirmation of a key prediction of the DSA model. 

The strength of the magnetic field in the shock acceleration region may be derived 
from the thickness of the nonthermal X-ray rims observed in young SNRs (e.g. 
\cite{vin03,bal06,par06}). One of the two interpretations that have been proposed to 
explain the thin X-ray filaments is that they result from fast synchrotron cooling 
of ultrarelativistic electrons transported downstream of the forward shock. In this 
scenario, the width of the filaments is set by the distance that the electrons cover 
before their synchrotron emission falls out of the X-ray band. The electron transport 
in the downstream region is due to a combination of diffusion and advection, whose 
corresponding scale heights are \cite{bal06} $l_{\rm diff}=
\sqrt{\kappa \tau_{\rm syn}}\propto B^{-3/2}$ (see eqs.[\ref{eqvt4}] and [\ref{eqvt6}]) 
and $l_{\rm adv}=\tau_{\rm syn}V_s/r_{\rm tot}\propto B^{-3/2} E_X^{-1/2} V_s/r_{\rm tot}$, 
respectively. Here, $r_{\rm tot}$ is the overall compression ratio of the shock and 
$E_X \sim 5$~keV is the typical X-ray energy at which the rims are observed. Thus, by 
comparing $l_{\rm diff}$ and $l_{\rm adv}$ to the measured width of the X-ray filaments 
(e.g. $l_{\rm obs}\approx 3''$ in Cas A which gives 0.05~pc for a distance of 3.4 kpc 
\cite{vin03}) one can estimate the downstream  magnetic field. Applications of this 
method to {\it Chandra} and {\it XMM-Newton} observations of young SNRs have shown that 
{\it the magnetic field at the forward shock is amplified by about two orders of 
magnitude} as compared with the average Galactic field strength. This conclusion has 
been recently strengthened by the observations of rapid time variations ($\sim$1~yr) in 
bright X-ray filaments of the SNR RXJ1713.7-3946 (also named G347.3-0.5), which are 
interpreted as resulting from fast synchrotron cooling of TeV electrons in a magnetic 
field amplified to milligauss levels \cite{uch07}. Such a high magnetic field is 
likely the result of a nonlinear amplification process associated with the efficient 
DSA of CRs \cite{bel01,ama06,vla06}. 

The other interpretation that has been proposed to account for the thin X-ray filaments 
is that they reflect the spatial distribution of the ambient magnetic field rather than
the spatial distribution of the ultrarealtivistic electrons \cite{poh05}. In this 
scenario, the magnetic field is thought to be amplified at the shock as well, but the 
width of the X-ray rims is not set by $l_{\rm diff}$ and $l_{\rm adv}$, but by the 
damping length of the magnetic field behind the shock. In this case, the relation given 
above between the rim thickness and the 
downstream magnetic field would not be valid. Comparison of high-resolution X-ray and 
radio images could allow to distinguish between the two interpretations, because 
the synchrotron energy losses are expected to be relatively small for GeV electrons 
emitting in the radio band \cite{vin03}. Thus, if the X-ray filaments are due to 
rapid synchrotron cooling of TeV electrons, the same structures should not be seen in 
radio images. A recent detailed study of Tycho's SNR has not allowed to draw firm 
conclusions on the role of magnetic damping behind the blast wave \cite{cas07}. 
Further high-resolution observations of SNRs in radio wavelengths would be very useful. 

The findings that (1) DSA can proceed at nearly the maximum possible rate (i.e. the Bohm 
limit) and (2) the magnetic field in the acceleration region can be strongly amplified, 
suggest that CR ions can reach higher energies in SNR shocks than previously estimated 
by Lagage \& Cesarsky \cite{lag83}. Thus, Berezhko \& V\"olk \cite{ber07} have argued 
that protons can be accelerated in SNRs up to the energy of the knee in the CR 
spectrum, at $3\times10^{15}$~eV. But relaxing the assumption of Bohm diffusion used 
in the calculations of Berezhko \& V\"olk, Parizot et al. \cite{par06} have obtained 
lower maximum proton energies for five young SNRs. These authors have derived an upper 
limit of $\sim8\times10^{14}$~eV on the maximum proton energy, $E_{p,\rm max}$, and 
have suggested that an additional CR component is required to explain the CR data 
above the knee energy. Recently, Ellison \& Vladimirov \cite{ell08} have pointed out 
that the average magnetic field that determines the maximum proton energy can be 
substantially less than the field that determines the maximum electron energy. This is 
because electrons remain in the vicinity of the shock where the magnetic field can be  
strongly amplified, whereas protons of energies $E_p>E_{e,\rm max}$ diffuse farther in 
the shock precursor region where the field is expected to be weaker 
($E_{p,\rm max}>E_{e,\rm max}$ because radiation losses affect the electrons only). 
This nonlinear effect of efficient DSA could reduce $E_{p,\rm max}$ relative 
to the value expected from test-particle acceleration. Nonetheless, recent calculations 
of $E_{p,\rm max}$ in the framework of nonlinear DSA models suggest that SNRs might 
well produce CRs up to the knee \cite{ell08,bla07}.

\section{TeV gamma-ray emission from SNRs}

Atmospheric Cerenkov telescopes have now observed high-energy $\gamma$ rays from six 
shell-type SNRs: Cas A with HEGRA \cite{aha01} and MAGIC \cite{alb07a}, 
RX~J1713.7-3946 with CANGAROO \cite{eno02} and HESS \cite{aha07a}, RX~J0852.0-4622 
(Vela Junior) with CANGAROO-II \cite{kat05} and HESS \cite{aha07b}, RCW~86 with HESS 
\cite{hop07}, IC~443 with MAGIC \cite{alb07b}, and very recently SN~1006 in deep HESS 
observations \cite{aha05b}. In addition, four $\gamma$-ray sources discovered in the 
Galactic plane survey performed with HESS are spatially coincident with SNRs \cite{aha06a}. 

With an angular resolution of $\sim0.06^\circ$ for individual $\gamma$ rays
\cite{aha07a,aha07b}, HESS has provided detailed images above 100 GeV of the extended 
SNRs RX~J1713.7-3946 and RX~J0852.0-4622 (their diameters are $\sim1^\circ$ and 
$\sim2^\circ$, respectively). In both cases the images show a shell-like structure 
and there is a striking correlation between the morphology of the $\gamma$-ray 
emission and the morphology previously
observed in X-rays. For both objects the X-ray emission is completely dominated by 
nonthermal synchrotron radiation. The similarity of the $\gamma$-ray and X-ray images 
thus suggest that the high-energy emission might also be produced by ultrarelativistic
electrons, via inverse Compton (IC) scattering off cosmic-microwave-background (CMB), 
optical-starlight and infrared photons. The $\gamma$-ray radiation would then be 
produced by electrons of energy (in the Thompson limit)
\begin{equation}
E_e \sim \bigg({3 \over 4}{E_\gamma \over E_\star}\bigg)^{1/2} m_e c^2,
\label{eqvt8}
\end{equation}
where $E_\star$ is the typical energy of the seed photons, $E_\gamma$ is the average 
final energy of the upscattered photons and $m_e$ is the electron  mass. For the CMB, 
whose contribution to the total IC emission of SNRs generally dominates, $E_\star \sim 
3kT_{\rm CMB}=7.1\times10^{-4}$~eV ($k$ is the Boltzmann constant and 
$T_{\rm CMB}=2.73$~K). Significant $\gamma$-ray emission beyond $E_\gamma=30$~TeV has 
been detected from RX~J1713.7-3946 \cite{aha07a}. Thus, an IC origin for the high-energy 
emission would imply that electrons are accelerated to more than 90~TeV in this object. 
In more accurate calculations that take into account the contributions of the optical 
and infrared interstellar radiation fields, the maximum electron energy is found to be 
$E_{e,\rm max} \sim 15$--40~TeV \cite{por06}. This result is consistent with the value 
of $E_{e,\rm max}$ derived from the width of X-ray filaments in RX~J1713.7-3946, 
$E_{e,\rm max}=36$~TeV \cite{par06}. 

Assuming that the same population of ultrarelativistic electrons produce both the 
observed TeV $\gamma$-rays and nonthermal X-rays, the mean magnetic field in the 
interaction region can be readily estimated from the ratio of synchrotron to IC 
luminosities:
\begin{equation}
{L_{\rm syn} \over L_{\rm IC}}={U_B \over U_{\rm rad}}={B^2 \over 8\pi U_{\rm rad}},
\label{eqvt9}
\end{equation}
where $U_B=B^2 / (8\pi)$ is the magnetic field energy density and $U_{\rm rad}$ is 
the total energy density of the seed photon field. With 
$U_{\rm CMB}\approx 0.25$~eV~cm$^{-3}$ for the CMB and 
$U_{\rm IR}\approx 0.05$~eV~cm$^{-3}$ for the interstellar infrared background (e.g. 
\cite{aha06b}), we have $U_{\rm rad}\approx 0.3$~eV~cm$^{-3}$ (we neglect here the 
contribution to the IC emission of the optical starlight background). Then, from the
measured ratio $L_{\rm syn} / L_{\rm IC}\approx 10$ for RX~J1713.7-3946 \cite{aha06b}, 
we obtain $B\approx 11~\mu$G. This value is significantly lower than the downstream 
magnetic field estimated from the observed X-ray rims: $B\sim80~\mu$G \cite{par06} (or 
$B>65~\mu$G in Ref.~\cite{ber06}). In other words, if the magnetic
field in the electron interaction region is as high as derived from the width of the 
X-ray filaments, IC radiation cannot account for the TeV $\gamma$-ray data. 

The magnetic field amplification is the main argument to favor a hadronic origin for 
the high-energy $\gamma$ rays produced in RX~J1713.7-3946 and other SNRs \cite{ber06}. 
The shape of the measured $\gamma$-ray spectrum below $\sim$1~TeV has also been used
to advocate that the high-energy emission might not be produced by IC scattering 
\cite{aha06b}, but the IC calculations of Ref.~\cite{por06} reproduce the broadband 
emission of RX~J1713.7-3946 reasonably well. In the hadronic scenario, the TeV 
$\gamma$ rays are due to nuclear collisions of accelerated protons and heavier 
particles with ambient ions, which produce neutral pions $\pi^0$ that decay in 99\% 
of the cases into two photons with energies of 67.5~GeV each in the $\pi^0$ rest 
frame ($2\times67.5$~GeV is the $\pi^0$ mass). At TeV energies in the observer rest 
frame, the spectrum of the $\pi^0$-decay $\gamma$ rays essentially reproduces,
with a constant scaling factor, the one of the parent ultrarelativistic particles. 
The accelerated proton energies can be estimated from the $\gamma$-ray spectrum 
as $E_p \sim E_\gamma/0.15$ \cite{aha07a}. The detection of $\gamma$ rays with 
$E_\gamma>30$~TeV in RX~J1713.7-3946 thus implies that protons are accelerated to more 
than 200~TeV, which is still about an order of magnitude below the energy of the knee. 

However, the hadronic scenario is problematic for RX~J1713.7-3946. Due to the lack of 
thermal X-ray emission, the remnant is thought to expand mostly in a very diluted 
medium of density $n<0.02$~cm$^{-3}$ \cite{cas04}. It is likely that the SN 
exploded in a bubble blown by the wind of the progenitor star. The flux of $\gamma$ 
rays produced by pion decay is proportional to the product of the number of 
accelerated protons and the ambient medium density. Thus, the total energy contained
in CR protons would have to be large to compensate the low density of the ambient 
medium. From the $\gamma$-ray flux measured with HESS from the center of the remnant, 
Plaga \cite{pla08} has recently estimated that the total CR-proton energy would have to 
be $>4\times10^{51}$~erg! Katz \& Waxman \cite{kat08} also argue against a hadronic 
origin for the TeV emission from RX~J1713.7-3946. They show that it would require that 
the CR electron-to-proton abundance ratio at a given relativistic energy 
$K_{\rm ep}\lsim2\times10^{-5}$, which is inconsistent with the limit they derived from 
radio observations of SNRs in the nearby galaxy M33, $K_{\rm ep}\gsim10^{-3}$. Moreover, radio 
and X-ray observations of RX~J1713.7-3946 suggest that the blast wave has recently hit 
a complex of molecular clouds located in the western part of the remnant \cite{cas04}. 
The ambient medium density in this region has been estimated to be $\sim300$~cm$^{-3}$. 
In the hadronic scenario, a much higher $\gamma$-ray flux would be expected in this 
direction, contrary to the observations \cite{pla08}. Thus, the $\gamma$-ray morphology 
revealed by HESS practically rules out pion production as the main contribution to 
the high-energy radiation of RX~J1713.7-3946.

But then, why the magnetic field given by the ratio of synchrotron to IC luminosities 
(eq.~[\ref{eqvt9}]) is inconsistent with the field derived from the X-ray filaments? 
This suggests that the filamentary structures observed with {\it Chandra} and 
{\it XMM-Newton} are localized regions where the magnetic field is enhanced in 
comparison with the mean downstream field \cite{kat08}. It is possible that the 
magnetic field is amplified at the shock as part of the nonlinear DSA process, but then 
rapidly damped behind the blast wave \cite{poh05}. The mean field downstream the shock
would then not be directly related to the observed thickness of the X-ray rims. 

Although the unambiguous interpretation of the TeV observations of shell-type SNRs 
remains uncertain, the high-energy $\gamma$-ray emissions from RX~J1713.7-3946 and 
RX~J0852.0-4622 are probably produced by IC scattering \cite{kat08}. For Cas~A, the case 
for a hadronic origin of the TeV radiation may be more compelling, as the density of the 
ambient medium is higher \cite{vin03}. The new source MAGIC~J0616+225 \cite{alb07b} which 
is spatially coincident with IC~443 may also be produced by pion decay. IC~443 is one of 
the best candidates for a $\gamma$-ray source produced by interactions between CRs 
accelerated in a SNR and a nearby molecular cloud \cite{tor03}. Hopefully the upcoming
{\it GLAST} satellite will allow a clear distinction between hadronic and electronic 
$\gamma$-ray processes in these objects. With the expected sensitivity of the LAT 
instrument between 30 MeV and 300 GeV, {\it GLAST} observations of SNRs should 
differentiate between pion-decay and IC spectra \cite{fun08}.

\section{Radio emission and nonlinear diffusive shock acceleration in SN 1993J}

About 30 extragalactic SNe have now been detected at radio wavelengths\footnote{See 
http://rsd-www.nrl.navy.mil/7213/weiler/kwdata/rsnhead.html.}. In a number of cases, 
the radio evolution has been monitored for years after outburst. It is generally 
accepted that the radio emission is nonthermal synchrotron radiation from relativistic 
electrons accelerated at the SN shock wave \cite{che82}. At early epochs the radio flux 
can be strongly attenuated by free-free absorption in the wind lost from the progenitor 
star prior to the explosion. Synchrotron self-absorption can also play a role in some 
objects \cite{che98}. The radio emission from SNe can provide unique information on 
the physical properties of the circumstellar medium (CSM) and the final stages of 
evolution of the presupernova system \cite{wei02}. We show here that this emission 
can also be used to study critical aspects of the DSA mechanism. 

The type IIb SN~1993J, which exploded in the nearby galaxy M81 at a distance of 
$3.63\pm0.34$~Mpc, is one of the brightest radio SNe ever detected (see \cite{wei07} 
and references therein). Very long baseline interferometry (VLBI) imaging has revealed 
a decelerating expansion of a shell-like radio source, which is consistent with the
standard model that the radio emission arises from a region behind the forward shock 
propagating into the CSM. The expansion has been found to be self-similar 
\cite{mar97}, although small departures from a self-similar evolution have been 
reported \cite{bar00}. The velocity of the forward shock can be estimated from the 
measured outer radius of the radio shell, i.e. the shock radius $r_s$, as 
$V_s=dr_s/dt=3.35\times10^4~t_d^{-0.17}$~km~s$^{-1}$, where $t_d$ is the time after 
shock breakout expressed in days. 

\begin{figure}
\includegraphics[width=1\textwidth]{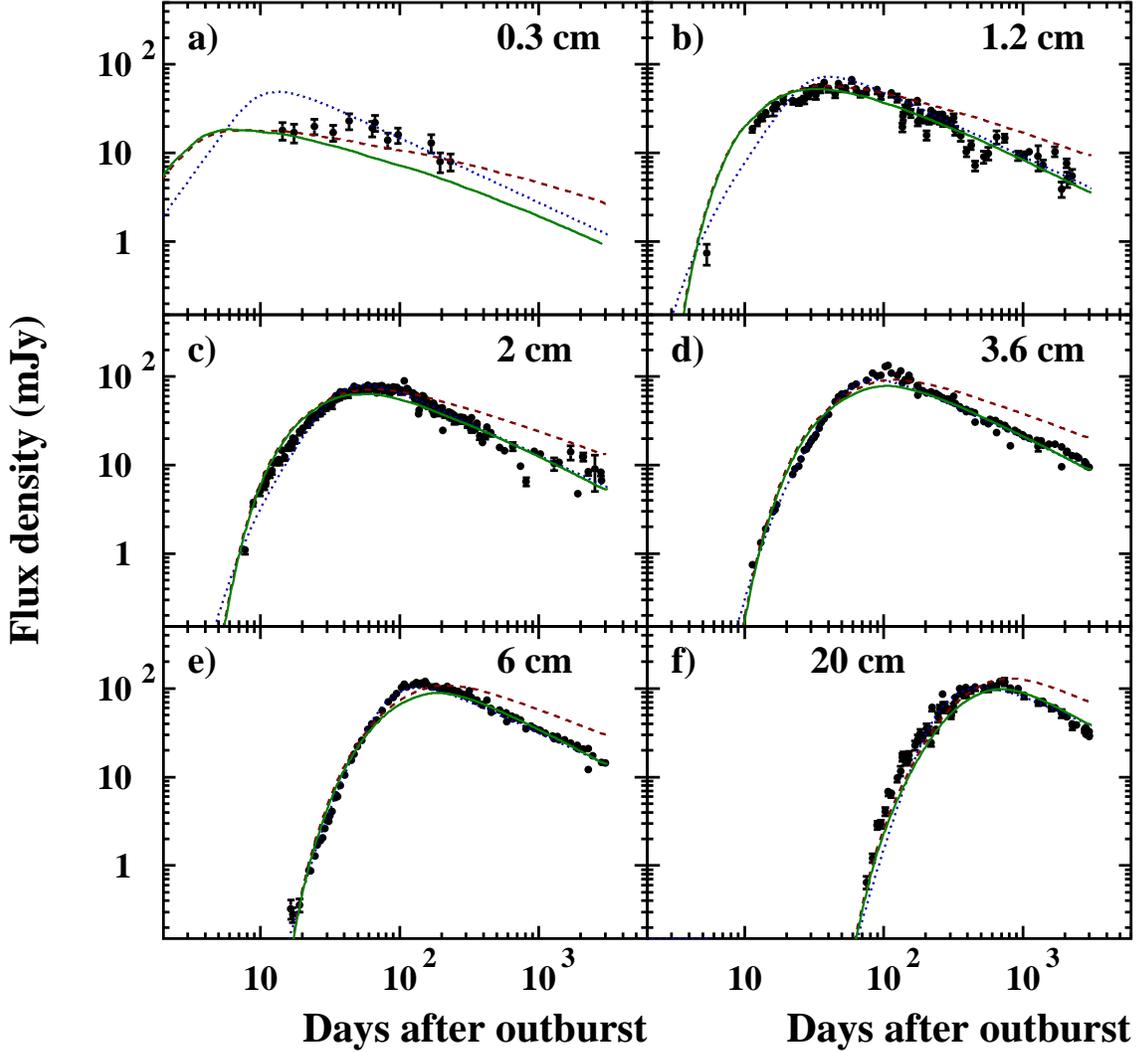}
\caption{Radio light curves for SN~1993J at 0.3, 1.2, 2, 3.6, 6, and 20~cm. The dotted
blue lines represent the best fit semi-empirical model of Ref.~\cite{wei07}. The 
dashed red (resp. solid green) lines show results of the present model for 
$\eta_{\rm inj}^p=\eta_{\rm inj}^e=10^{-5}$ (resp. 
$\eta_{\rm inj}^p=2\times10^{-4}$ and $\eta_{\rm inj}^e=1.4\times10^{-5}$; see text).
The data are from Ref.~\cite{wei07} and references therein. 
}
\label{figvt1}
\end{figure}

Extensive radio monitoring of the integrated flux density of SN~1993J has been 
conducted with the Very Large Array and several other radio telescopes \cite{wei07}. 
Figure~\ref{figvt1} shows a set of measured light curves at 0.3~cm (85--110 GHz), 
1.2~cm (22.5~GHz), 2~cm (14.9~GHz), 3.6~cm (8.4~GHz), 6~cm (4.9~GHz), and 20~cm 
(1.4~GHz). We see that at each wavelength the flux density first rapidly increases 
and then declines more slowly as a power in time (the data at 0.3~cm do not allow to 
clearly identify this behavior). The radio emission was observed to suddenly decline 
after day $\sim$3100 (not shown in Fig.~\ref{figvt1}), which is interpreted in terms 
of an abrupt decrease of the CSM density at radial distance from the progenitor 
$r\sim3\times10^{17}$~cm \cite{wei07}. The maximum intensity is reached first at lower 
wavelengths and later at higher wavelengths, which is characteristic of absorption 
processes. For SN~1993J, both free-free absorption in the CSM and synchrotron 
self-absorption are important \cite{che98,fra98,wei07}. To model light curves of radio 
SNe, Weiler et al. \cite{wei02,wei07} have developed a semi-empirical formula that 
takes into account these two absorption mechanisms. For SN~1993J, the best fit to the 
data using this semi-empirical model ({\it dotted blue curves} in Fig.~\ref{figvt1}) 
requires nine free parameters \cite{wei07}. 

I have developed a model for the radio emission of SN~1993J, which is inspired by 
previous works on the morphology of synchrotron emission in young SNRs 
\cite{cas05,ell05}. The model will be presented in detail in a forthcoming publication 
\cite{tat08} and I only give here broad outlines. First, the density profile for the 
CSM is taken as $\rho_{\rm CSM}(r)=\rho_0(r/r_0)^{-2}$ as expected for a constant wind 
mass-loss rate and terminal velocity. Here $r_0=3.49\times10^{14}$~cm is the shock 
radius at $t=1$~day after outburst and $\rho_0$ is a free parameter. Evidence for a 
flatter CSM density profile has been advocated ($\rho_{\rm CSM}\propto r^{-s}$, 
with $s \sim 1.6$; see \cite{wei07} and references therein), based on the measured time 
dependence of the optical depth to free-free absorption in the CSM, $\tau_{\rm ff}$. 
However, Fransson \& Bj\"ornsson \cite{fra98} have shown that the time 
evolution of $\tau_{\rm ff}$ can be explained by a decrease of the CSM temperature with 
$r$ together with the standard $r^{-2}$ distribution for the density. The results of the
present work provide support to this latter interpretation \cite{tat08}. Thus, in the 
present model, free-free absorption is calculated assuming the time dependence of 
$\tau_{\rm ff}$ obtained in Ref.~\cite{fra98} and using the best-fit value of $\rho_0$ 
(or more precisely $\rho_0^2$) as a normalization factor. 

Because synchrotron self-absorption is important in SN~1993J, the strength and evolution 
of the mean magnetic field in the region of the radio emission can be estimated from 
the measured peak flux at different wavelengths \cite{che98}. Using equation~(12) of 
Ref.~\cite{che98}, I obtain from the data at 1.2, 2, 3.6, 6, and 20~cm: 
\begin{equation}
\langle B \rangle = (46 \pm 19) \alpha^{-2/9} t_d^{-1.01\pm0.09}~{\rm G}, 
\label{eqvt10}
\end{equation}
where $\alpha$ is the ratio of the total energy density in relativistic electrons to
the magnetic energy density. The errors include the uncertainty in the contribution of 
free-free absorption. This time dependence of $\langle B \rangle$ is close to that 
expected if the magnetic field at the shock is amplified by a constant factor from 
the available kinetic energy density
(see \cite{bel01}). In this case, one expects $B^2/8\pi \propto \rho_{\rm CSM} V_s^2$,
which gives $B\propto t^{-1}$ for $\rho_{\rm CSM}\propto r^{-2}$. Note that the 
flatter CSM density profile supported by Weiler et al. \cite{wei07},  
$\rho_{\rm CSM}\propto r^{-1.6}$, would imply for the assumed scaling 
$B\propto \rho_{\rm CSM}^{1/2} V_s \propto t^{-0.83}$, which is somewhat inconsistent 
with the data. 

Based on the measured time dependence of $\langle B \rangle$, the immediate postshock 
magnetic field is assumed to be of the form $B_d=B_0t_d^{-1}$, where $B_0$ is a free
parameter expected to be in the range $\sim$100--600~G for a typical value of $\alpha$ 
in the range $\sim$10$^{-5}$--10$^{-2}$ (see eq.~[\ref{eqvt10}]). The evolution of 
the magnetic field behind the shock is then calculated from the assumption that the 
field is carried by the flow, frozen in the plasma, so that the parallel and 
perpendicular magnetic field components evolve conserving flux (see \cite{cas05} and 
references therein). Results obtained with the alternative assumption that the 
magnetic field is rapidly damped behind the shock wave will be given in \cite{tat08}. 

The hydrodynamic evolution of the plasma downstream the forward shock is calculated 
from the two-fluid, self-similar model of Chevalier \cite{che83}, which takes into 
account the effects of CR pressure on the dynamics of 
the thermal gas. The overall structure of SNRs can be described by self-similar 
solutions, if the initial density profiles in the ejected material (ejecta) 
and in the ambient medium have power-law distributions, and if the ratio of 
relativistic CR pressure to total pressure at the shock front is constant \cite{che83}. 
The backreaction of energetic ions can strongly modify the shock structure of young 
SNRs, such as e.g. Kepler's remnant \cite{dec00}. But the situation is different for 
SN~1993J because of the much higher magnetic field in the shock precursor region, 
which implies that energy is very efficiently transfered from the CRs to the thermal 
gas via Alfv\'en wave dissipation \cite{ber99}. The resulting increase in the gas 
pressure ahead of the viscous subshock is found to limit the overall compression ratio, 
$r_{\rm tot}$, to values close to 4 (i.e. the standard value for a test-particle 
strong shock) even for efficient DSA. Thus, the hydrodynamic evolution of SN~1993J 
can be safely calculated in the test-particle, self-similar approximation. 

Both the energy spectra of the accelerated particles and the thermodynamic properties of 
the gas just behind the shock front (i.e. the boundary conditions for the self-similar 
solutions of the hydrodynamic evolution) are calculated with the semianalytic model of 
nonlinear DSA developed by Berezhko \& Ellison \cite{ber99} and Ellison et al. 
\cite{ell00}. However, a small change to the model has been made: the Alfv\'en waves 
are assumed to propagate isotropically in the precursor region and not only in the 
direction opposite to the plasma flow (i.e. eqs.~(52) and (53) of Ref.~\cite{ber99} are 
not used). This is a reasonable assumption given the strong, nonlinear magnetic field 
amplification \cite{bel01}. Although the semianalytic model strictly applies to 
plane-parallel, steady state shocks, it has been sucessfully used in Ref.~\cite{ell00} 
for evolving SNRs. The main parameter of this model, 
$\eta_{\rm inj}^p$, is the fraction of total shocked protons in protons with momentum 
$p \geq p_{\rm inj}$ injected from the postshock thermal pool into the DSA process. 
The work of Ref.~\cite{bla05} allows us to accurately relate the injection momentum 
$p_{\rm inj}$ to $\eta_{\rm inj}^p$. Similarly, we define $\eta_{\rm inj}^e$ for the 
electron injection. The latter parameter is not important for the shock structure, 
because the fraction of total particle momentum carried by electrons is negligible, but 
it determines the brightness of optically-thin synchrotron emission for a given magnetic 
field strength. The electrons accelerated at the shock experience adiabatic and 
synchrotron energy losses as they are advected downstream with the plasma flow. The 
spectral evolution caused by these losses is calculated as in Ref.~\cite{rey98}. 
Finally, once both the nonthermal electron distribution and the magnetic field in a 
given shell of material behind the forward shock have been determined, the synchrotron 
emission from that shell can be calculated \cite{pac70}. The total radio emission along 
the line of sight is then obtained from full radiative transfer calculations that 
include synchrotron self-absorption. 

With the set of assumptions given above, the model has four free parameters: $\rho_0$, 
$B_0$, $\eta_{\rm inj}^p$ and $\eta_{\rm inj}^e$. While the product $\rho_0 \times 
\eta_{\rm inj}^e$ is important for the intensity of optically-thin synchrotron 
emission, the CSM density normalization $\rho_0$ also determines the level of free-free
absorption. The main effect of changing $B_0$ is to shift the radio light curves in time,
because the turn-on from optically-thick to optically-thin synchrotron emission is 
delayed when the magnetic field is increased. The proton injection paramater 
$\eta_{\rm inj}^p$ determines the shock structure and hence influences the shape of the
electron spectrum. 

Calculated radio light curves are shown in Fig.~\ref{figvt1} for 
$\rho_0=1.8\times10^{-15}$~g~cm$^{-3}$, $B_0=400$~G, and two sets of injection 
parameters: $\eta_{\rm inj}^p=\eta_{\rm inj}^e=10^{-5}$ (test-particle case), and 
$\eta_{\rm inj}^p=2\times10^{-4}$ and $\eta_{\rm inj}^e=1.4\times10^{-5}$. We see that 
in the test-particle case, the decline of the optically-thin emission with time is too
slow as compared to the data, except at 0.3~cm. The CR-modified shock provides a better 
overall fit to the measured flux densities, although significant deviations from the 
data can be observed. In particular, we see that the calculated light curve at 0.3~cm 
falls short of the data at this wavelength. It is possible that the
deviations of the best-fit curves from the data partly arise from the 
approximations used in the DSA model of Ref.~\cite{ber99}, in which the nonthermal 
phase-space distribution function $f(p)$ is described as a three-component power law. 
But it is also possible that it tells us something about the magnetic field evolution 
in the downstream region. The spatial distribution of the postshock magnetic field will 
be studied in Ref.~\cite{tat08} by comparing calculated synchrotron profiles with the 
observed average profile of the radio shell. 

\begin{figure}
\includegraphics[width=1\textwidth]{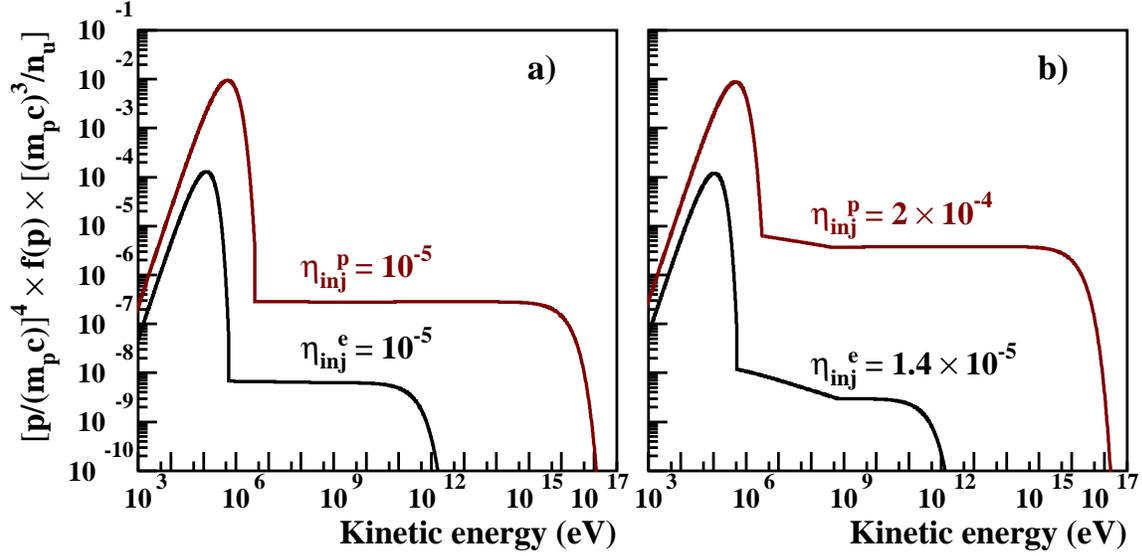}
\caption{Calculated postshock phase space distribution functions, $f(p)$, vs. kinetic 
energy, at day 100 after shock breakout. Following Ref.~\cite{ber99}, $f(p)$ 
has been multiplied by $[p/(m_pc)]^4$ to flatten the spectra, and by $[(m_pc)^3/n_u]$ to 
make them dimensionless ($m_p$ is the proton mass and $n_u$ the proton number density 
ahead of the shock precursor). The red lines are for protons and the black lines for 
electrons. The two sets of injection parameters $\eta_{\rm inj}^p$ and $\eta_{\rm inj}^e$ 
are those used for the synchrotron calculations shown in Fig.~1.}
\label{figvt2}
\end{figure}

For SN~1993J, the main effect of the CR pressure is to reduce the compression ratio 
of the subshock, $r_{\rm sub}$, whereas the overall compression ratio $r_{\rm tot}$ 
remains nearly constant (see above). For $\eta_{\rm inj}^p=2\times10^{-4}$, $r_{\rm sub}$ 
is found to decrease from 3.58 to 3.35 between day 10 and day 3100 after outburst, whereas 
$r_{\rm tot}$ stays between 4 and 4.04. Such a shock modification affects essentially 
the particles of energies $<m_pc^2$ that remain in the vicinity of the subshock during 
the DSA process. Thus, we see in Fig.~2 that the increase of $\eta_{\rm inj}^p$ from 
10$^{-5}$ to $2\times10^{-4}$ steepens the phase-space distribution functions $f(p)$
of both protons and electrons between their thermal Maxwell-Boltzmann distributions and 
$\sim1$~GeV, as the spectral index in this energy domain,  
$q_{\rm sub}=3r_{\rm sub}/(r_{\rm sub}-1)$ with $f(p)\propto p^{-q_{\rm sub}}$ 
\cite{ber99}, increases with decreasing $r_{\rm sub}$. The radio light curves provide 
clear evidence that the spectral index of the nonthermal electrons below 1~GeV is higher 
than the strong-shock test-particle value $q_{\rm sub}=4$. For the preferred injection 
parameters $\eta_{\rm inj}^p=2\times10^{-4}$ and $\eta_{\rm inj}^e=1.4\times10^{-5}$, 
the calculated electron-to-proton ratio at, e.g., 10~GeV is $K_{\rm ep}=7.8\times10^{-4}$ 
at $t_d=100$~days (Fig.~2) and $K_{\rm ep}=6.3\times10^{-4}$ at $t_d=3100$~days. These values 
are roughly consistent with -- but slightly lower than -- those recently estimated for the 
blast wave of Tycho's SNR, $K_{\rm ep}\sim10^{-3}$ \cite{cas07}. 

\begin{figure}
\begin{center}
\includegraphics[width=.7\textwidth]{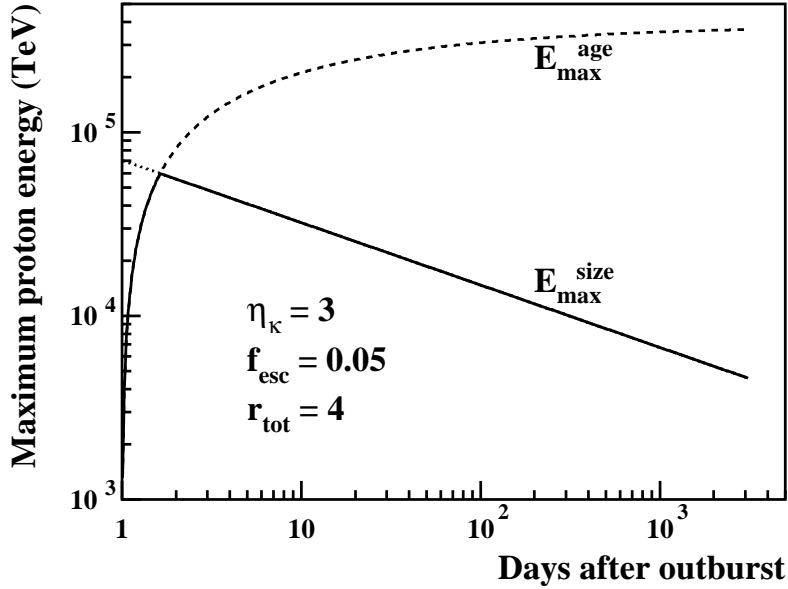}
\end{center}
\caption{Estimated maximum proton energy in SN~1993J as a function of time after shock 
breakout. $E_{\rm max}^{\rm age}$ and $E_{\rm max}^{\rm size}$ are the maximum energies 
caused by the finite shock age and size, respectively. The resulting maximum proton 
energy ({\it solid line}) is the minimum of these two quantities.}
\label{figvt3}
\end{figure}

The immediate postshock magnetic field obtained in this work, $B_d\approx 400\times 
t_d^{-1}$~G is in very good agreement with the mean field in the synchrotron-emitting 
region previously estimated by Fransson \& Bj\"ornsson \cite{fra98}, 
$\langle B \rangle\approx 370\times t_d^{-1}$~G. As already pointed out by Bell \& Lucek 
\cite{bel01}, the derived magnetic field strength in the forward shock precursor is 
consistent with a simple estimate based on nonlinear magnetic field 
amplification driven by the pressure gradient of accelerated particles. For 
$\eta_{\rm inj}^p=2\times10^{-4}$, the ratio of the mean magnetic energy density in the 
shock precursor to the energy density in CR protons is found to slightly decrease from 0.6 
to 0.5 between day 10 and day 3100 after outburst \cite{tat08}. It is interesting that 
the magnetic field strength is very close to that given by equipartition.  

As also pointed out in Ref.~\cite{bel01}, CRs were probably accelerated to very high 
energies in SN~1993J short after shock breakout. The maximum proton energy $E_{p,\rm max}$ 
is expected to be limited by the spatial extend of the shock, because high-energy 
particles diffusing upstream far enough from the shock front can escape from the 
acceleration region. This size limitation can be estimated by assuming the existence of 
an upstream free escape boundary located at some constant fraction $f_{\rm esc}$ of the 
shock radius: $d_{\rm FEB}=f_{\rm esc}r_s$. The maximum energy that particles can acquire 
before reaching this boundary is then obtained by equalling $d_{\rm FEB}$ to the upstream 
diffusion length, $l \sim \kappa / V_s$ (e.g. \cite{ell08}). Calculated time evolution of 
$E_{p,\rm max}$ is shown in Fig.~\ref{figvt3} for $f_{\rm esc}=0.05$ (see, e.g., \cite{ell05}) 
and $\eta_\kappa=3$ (see eq.~\ref{eqvt7} and, e.g., \cite{par06}). We see that the maximum 
energy caused by the particle  escape, $E_{\rm max}^{\rm size}$ 
($\propto f_{\rm esc}\eta_\kappa^{-1}$), becomes rapidly lower than the maximum energy 
associated with the shock age, $E_{\rm max}^{\rm age}$ ($\propto \eta_\kappa^{-1}$), which 
is obtained by time integration of the acceleration rate $(dE/dt)_{\rm acc}$ from an initial 
acceleration time assumed to be $t_0=1$~day after outburst. The resulting maximum proton 
energy (i.e. the minimum of $E_{\rm max}^{\rm size}$ and $E_{\rm max}^{\rm age}$) is 
significantly lower than the previous estimate of Ref.~\cite{bel01}, 
$E_{p,\rm max}\sim3\times10^{17}$~eV. Moreover, we did not take into account the nonlinear 
effects recently pointed out by Ellison \& Vladimirov \cite{ell08}, which could further 
reduce $E_{p,\rm max}$. However, SN~1993J may well have accelerated protons above the knee 
energy of $3\times10^{15}$~eV (Fig.~\ref{figvt3}), which provides support to the scenario 
first proposed by V\"olk \& Biermann \cite{vol88} that the highest-energy Galactic CRs are 
produced by massive stars exploding into their own wind. In this context, it is instructive 
to estimate the total energy acquired by the CR particles during the early phase of 
interaction between the SN ejecta and the red supergiant wind:
\begin{equation}
E_{\rm CR} \cong \int_{t_d=1}^{3100} \epsilon_{\rm CR}(t) \times {1 \over 2} 
\rho_{\rm CSM}(r_s) V_s^3 \times 4\pi r_s^2 dt = 6.8 \times 10^{49}~\rm erg,
\label{eqvt11}
\end{equation}
where $\epsilon_{\rm CR}(t)$ is the time-dependent fraction of total incoming energy 
flux, $F_0(r_s) \cong 0.5\rho_{\rm CSM}(r_s) V_s^3$, going into CR particles. With the 
best-fit parameters obtained from the radio light curves, $\epsilon_{\rm CR}(t)$ is found 
to slowly increase from 12\% to 16\% between day 10 and day 3100 after outburst. It is 
remarkable that the value obtained for $E_{\rm CR}$ is very close to the mean energy per 
SN required to account for the Galactic CR luminosity, $\approx 7.5 \times 10^{49}$~erg 
(see Sect.~1), which suggests that SN~1993J might be typical of the SNe that produce the 
CR population in our own Galaxy. 

\section{Conclusions}

Observations of young shell-type SNRs with the {\it Chandra} and {\it XMM-Newton} X-ray 
space observatories give more credit to the current paradigm that the bulk of Galactic 
CRs are accelerated in shock waves generated by SN explosions. Evidence is accumulating
suggesting that acceleration in SN shocks can be nearly as fast as the Bohm limit and 
that self-excited turbulence in the shock vicinity can strongly amplify the ambient 
magnetic field, thus allowing the acceleration of CR ions to energies above 10$^{15}$~eV. 
Besides, observations of nonlinear effects caused by efficient DSA support the fact that 
the acceleration efficiency can be high enough to account for the Galactic CR luminosity. 
But the direct and unambiguous observation of ion acceleration in SN shocks 
remains uncertain, as it is difficult at the present time to distinguish between 
pion decay and IC scattering as the main mechanism responsable for the very-high-energy 
$\gamma$-ray emission detected with ground-based atmospheric Cerenkov telescopes from 
several shell-type SNRs. Following Refs.~\cite{pla08,kat08} I have argued, however, that 
the TeV $\gamma$ rays emitted in RX~J1713.7-3946 and RX~J0852.0-4622 are probably 
produced by ultrarelativistic electrons, via IC scattering off CMB, optical-starlight and 
infrared photons. One of the most critical ingredients for the interpretation of these 
observations is the evolution of the postshock magnetic field in the downstream region 
(see Sect.~3). Hopefully, the upcoming GLAST satellite (launch scheduled for May 2008) 
will allow a clear identification of the contributions of hadronic and electronic 
$\gamma$-ray emission processes in SNRs, which in turn will be useful to understand the 
importance of magnetic field damping behind the forward shock.

Radio observations of extragalactic SNe can also shed light on the DSA process. The 
impressive set of radio data available for SN~1993J make this 
object a unique laboratory to study particle acceleration in a SN shock. Using a 
semianalytic model of nonlinear DSA to explain the radio light curves, the results I have 
obtained suggest, in particular, that the acceleration of ions has become efficient soon 
after outburst and that the magnetic field in the forward shock vicinity has been amplified 
to equipartition with the CR energy density. The field amplification implies that CR 
protons may have been accelerated to energies above $3\times10^{15}$~eV during the early 
phase of interaction between the SN ejecta and the red supergiant wind lost from the 
progenitor star. During this time, which lasted only $\sim$8.5~years, the shock processed 
in the expansion a total energy of $4.6\times10^{50}$~erg and the mean acceleration 
efficiency was found to be $\epsilon_{\rm CR}=15\%$. Thus, a total energy of almost 
$7\times10^{49}$~erg has been stored up by CR particles during this phase. A significant 
fraction of this energy might have escaped into the interstellar medium after day 
$\sim$3100, when the shock started to expand into a more diluted CSM. 

The results obtained for SN~1993J suggest that massive stars exploding into a wind 
environment could be a major source of Galactic CRs, as first proposed by V\"olk 
\& Biermann \cite{vol88}. It is well known that most massive stars are born in OB 
associations, whose activity formes superbubbles of hot and turbulent gas inside 
which the majority of core-collapse SNe explode. Turbulent re-acceleration inside 
superbubbles may modify the spectrum of CRs and, in particular, increase their 
maximum energy (see \cite{par04} and references therein). 

\acknowledgements{It is a pleasure to thank Anne Decourchelle and J\"urgen Kiener for 
critical reading of the manuscipt, Margarita Hernanz for numerous discussions and her 
hospitality at IEEC-CSIC, and Jordi Isern for the invitation to the conference. 
Financial support from the Generalitat de Catalunya through the AGAUR grant 
2006-PIV-10044 and the project SGR00378 is acknowledged.}


\begin{thebibliography}{99}
\bibitem{sre93} P. Sreekumar et al., \emph{Phys. Rev. Lett.} {\bf 70} (1993) 127
\bibitem{dog02} V.~A. Dogiel, V. Sch{\"o}nfelder, \& A.~W. Strong, 
\emph{Astrophys. J.} {\bf 572} (2002) L157  
\bibitem{fer98} K. Ferri\`ere, \emph{Astrophys. J.} {\bf 497} (1998) 759
\bibitem{kry77} G. F. Krymskii, \emph{Dokl. Akad. Nauk SSSR} {\bf 234} (1977) 1306
\bibitem{axf78} W. I. Axford, E. Leer, \& G. Skadron, in \emph{Proc. 15th Int. 
Cosmic-Ray Conf.} ( Plovdiv) {\bf 11} (1978) 132 
\bibitem{bel78} A. R. Bell, \emph{Mon. Not. R. Astron. Soc.} {\bf 182} (1978) 147
\bibitem{bla78} R. D. Blandford \& J. Ostriker, \emph{Astrophys. J.} {\bf 221} 
(1978) L29
\bibitem{lag83} P. O. Lagage \& C. J. Cesarsky, \emph{Astron. Astrophys.}  
{\bf 125} (1983) 249
\bibitem{bel01} A. R. Bell \& S. G. Lucek, \emph{Mon. Not. R. Astron. Soc.} 
{\bf 327} (2001) 433
\bibitem{ama06} E. Amato \& P. Blasi, \emph{Mon. Not. R. Astron. Soc.} 
{\bf 371} (2006) 1251
\bibitem{vla06} A. Vladimirov, D. C. Ellison, \& A. Bykov \emph{Astrophys. J.} 
{\bf 652} (2006) 1246
\bibitem{ber07} E. G. Berezhko \& H. J. V\"olk, \emph{Astrophys. J.} {\bf 661} 
(2007) L175
\bibitem{bla05} P. Blasi, S. Gabici, \& G. Vannoni, \emph{Mon. Not. R. Astron. Soc.} 
{\bf 361} (2005) 907
\bibitem{ber99} E. G. Berezhko \& D. C. Ellison, \emph{Astrophys. J.} {\bf 526} 
(1999) 385
\bibitem{dec00} A. Decourchelle, D. C. Ellison, \& J. Ballet \emph{Astrophys. J.} 
{\bf 543} (2000) L57
\bibitem{hug00} J. P. Hughes, C. E. Rakowski, \& A. Decourchelle, 
\emph{Astrophys. J.} {\bf 543} (2000) L61
\bibitem{dec05} A. Decourchelle, in \emph{X-Ray and Radio Connections} (eds. 
L.O. Sjouwerman and K.K. Dyer), Published electronically by NRAO, 
http://www.aoc.nrao.edu/events/xraydio, 2005
\bibitem{war05} J. S. Warren et al., \emph{Astrophys. J.} {\bf 634} (2005) 376
\bibitem{koy95} K. Koyama et al., \emph{Nature} {\bf 378} (1995) 255
\bibitem{wei07} K. Weiler et al.,  \emph{Astrophys. J.} {\bf 671} (2007) 1959
\bibitem{cas04} G. Cassam-Chena{\"i} et al., \emph{Astron. Astrophys.} {\bf 427} 
(2004) 199
\bibitem{sta06} M. D. Stage, G. E. Allen, J. C. Houck, \& J. E. Davis, 
\emph{Nature Phys.} {\bf 2} (2006) 614
\bibitem{par06} E. Parizot, A. Marcowith, J. Ballet, \& Y. A. Gallant, 
\emph{Astron. Astrophys.} {\bf 453} (2006) 387
\bibitem{rey04} S. P. Reynolds, \emph{Adv. Space Res.} {\bf  33} (2004) 461
\bibitem{yam04} R. Yamazaki, T. Yoshida, T. Terasawa, A. Bamba, \& K. Koyama, 
\emph{Astron. Astrophys.} {\bf 416} (2004) 595
\bibitem{vin03} J. Vink \& J. M. Laming, \emph{Astrophys. J.} {\bf 584} (2003) 758
\bibitem{bal06} J. Ballet, \emph{Adv. Space Res.} {\bf  37} (2006) 1902
\bibitem{uch07} Y. Uchiyama, F. A. Aharonian, T. Tanaka, T. Takahashi, \& Y. Maeda, 
\emph{Nature} {\bf  449} (2007) 576
\bibitem{poh05} M. Pohl, H. Yan, \& A. Lazarian, \emph{Astrophys. J.} {\bf 626} 
(2005) L101 
\bibitem{cas07} G. Cassam-Chena\"i, J. P. Hughes, J. Ballet, \& A. Decourchelle, 
\emph{Astrophys. J.} {\bf 665} (2007) 315
\bibitem{ell08} D. C. Ellison \& A. Vladimirov, \emph{Astrophys. J.} {\bf 673} 
(2008) L47
\bibitem{bla07} P. Blasi, E. Amato, \& D. Caprioli, \emph{Mon. Not. R. Astron. Soc.} 
{\bf 375} (2007) 1471
\bibitem{aha01} F. Aharonian et al., \emph{Astron. Astrophys.} {\bf 370} (2001) 112
\bibitem{alb07a} J. Albert et al., \emph{Astron. Astrophys.} {\bf 474} (2007) 937
\bibitem{eno02} R. Enomoto et al., \emph{Nature} {\bf 416} (2002) 823
\bibitem{aha07a} F. Aharonian et al., \emph{Astron. Astrophys.} {\bf 464} (2007) 235
\bibitem{kat05} H. Katagiri et al., \emph{Astrophys. J.} {\bf 619} (2005) L163
\bibitem{aha07b} F. Aharonian et al., \emph{Astrophys. J.} {\bf 661} (2007) 236
\bibitem{hop07} S. Hoppe \& M. Lemoine-Goumard for the HESS collaboration, in
\emph{Proc. 30th Int. Cosmic-Ray Conf.}, in press, arXiv:0709.4103 
\bibitem{alb07b} J. Albert et al., \emph{Astrophys. J.} {\bf 664} (2007) L87
\bibitem{aha05b} F. Aharonian et al., in preparation
\bibitem{aha06a} F. Aharonian et al., \emph{Astrophys. J.} {\bf 636} (2006) 777
\bibitem{por06} T. A. Porter, I. V. Moskalenko, \& A. W. Strong, \emph{Astrophys. J.} 
{\bf 648} (2006) L29.
\bibitem{aha06b} F. Aharonian et al., \emph{Astron. Astrophys.} {\bf 449} (2006) 223
\bibitem{ber06} E. G. Berezhko \& H. J. V\"olk, \emph{Astron. Astrophys.} {\bf 451} 
(2006) 981
\bibitem{pla08} R. Plaga, \emph{New Astron.} {\bf 13} (2008) 73
\bibitem{kat08} B. Katz \& E. Waxman, \emph{J. Cosmology Astropart. Phys.} {\bf 01} 
(2008) 18
\bibitem{tor03} D. F. Torres, G. E. Romero, T. M. Dame, J. A. Combi, \& Y. M. Butt, 
\emph{Phys. Rep.} {\bf 382} (2003) 303
\bibitem{fun08} S. Funk, \emph{Adv. Space Res.} {\bf 41} (2008) 464
\bibitem{che82} R. A. Chevalier, \emph{Astrophys. J.} {\bf 259} (1982) 302
\bibitem{che98} R. A. Chevalier, \emph{Astrophys. J.} {\bf 499} (1998) 810
\bibitem{wei02} K. W. Weiler, N. Panagia, M. J. Montes, \& R. A. Sramek, \emph{Annu. 
Rev. Astron. Astrophys.} {\bf 40} (2002) 387
\bibitem{mar97} J. M. Marcaide et al., \emph{Astrophys. J.} {\bf 486} (1997) L31
\bibitem{bar00} N. Bartel et al., \emph{Science} {\bf 287} (2000) 112
\bibitem{fra98} C. Fransson \& C.-I. Bj\"ornsson, \emph{Astrophys. J.} {\bf 509} 
(1998) 861
\bibitem{cas05} G. Cassam-Chena\"i, A. Decourchelle, J. Ballet, \& D. C. Ellison,
\emph{Astron. Astrophys.} {\bf 443} (2005) 955
\bibitem{ell05} D. C. Ellison \& G. Cassam-Chena\"i, \emph{Astrophys. J.} {\bf 632} 
(2005) 920
\bibitem{tat08} V. Tatischeff, in preparation
\bibitem{che83} R. A. Chevalier, \emph{Astrophys. J.} {\bf 272} (1983) 765
\bibitem{ell00} D. C. Ellison, E. G. Berezhko, \& M. G. Baring, \emph{Astrophys. J.} 
{\bf 540} (2000) 292
\bibitem{rey98} S. P. Reynolds, \emph{Astrophys. J.} {\bf 493} (1998) 375
\bibitem{pac70} A. G. Pacholczyk, \emph{Radio Astrophysics}, San Francisco: Freeman, 
1970
\bibitem{vol88} H. J. V\"olk \& P. L. Biermann, \emph{Astrophys. J.} {\bf 333} (1988) 
L65
\bibitem{par04} E. Parizot, A. Marcowith, E. van der Swaluw, A. M. Bykov, and V. 
Tatischeff, \emph{Astron. Astrophys.} {\bf 424} (2004) 747
\end{thebibliography}
\end{document}